\documentclass{article}

\usepackage[utf8]{inputenc}
\usepackage{graphicx}
\usepackage{amssymb}
\usepackage{amsmath}
\usepackage{hyperref}
\usepackage{listings}
\usepackage{float}
\usepackage[T1]{fontenc}
\usepackage{pgfplots}
\pgfplotsset{compat=newest}
\usepackage{microtype}

\begin{document}

\begin{titlepage}
    \centering
    \vspace*{2cm}
    {\huge\bfseries PureLottery: Fair and Bias-Resistant Leader Election with a Novel Single-Elimination Tournament Algorithm\\}

    \vspace{1cm}
    {\Large Bachelor's Thesis in Informatics\\}
    \vspace{1cm}
    {\Large Jonas Ballweg\\}

    \vspace{5cm}
    
    \includegraphics[width=4cm]{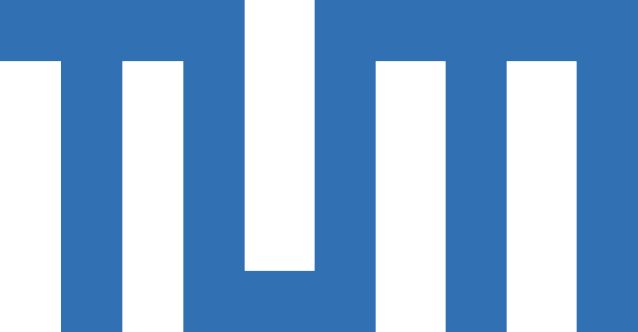} \\
    \vspace{0.5cm}
    {\Large\scshape Technical University of Munich\\}
    {\Large\scshape School of Computation, Information and Technology - Informatics\\} 
\end{titlepage}

\begin{titlepage}
    \centering
    {\huge\bfseries PureLottery: Fair and Bias-Resistant Leader Election with a Novel Elimination Tournament Algorithm\\}
    \vspace{1cm}
    {\huge\bfseries PureLottery: Faire und verzerrungsresistente Leiter-Wahl mit einem neuen Algorithmus für Ausscheidungsturniere\\}
    
    \vspace{1.5cm}
    {\Large Bachelor's Thesis in Informatics\\}
    \vspace{0.5cm}
    {\Large by Jonas Ballweg\\}
    \vspace{1cm}
    {\Large Examiner: Prof. Dr. Javier Esparza\\}
    {\Large Supervisor: Prof. Dr. Amir Goharshady\\}
    \vspace{0.5cm}
    {\Large Submission date: 2024-03-01}

    \vspace{3cm}
    {\Large\scshape Technical University of Munich\\}
    {\Large\scshape School of Computation, Information and Technology - Informatics\\}
\end{titlepage}

\newpage
\noindent I confirm that this bachelor's thesis is my own work and I have documented all sources and material used. \\
\vspace{1.2cm} \\
\rule{8cm}{0.4pt}

\newpage
\noindent \textbf{\Large Acknowledgements}
\vspace{0.3cm}
\\
\noindent My deep gratitude is owed to Professor Amir Goharshady, not only for his expert supervision throughout this thesis but also for his mentorship in authoring my first academic paper. I want to thank him for introducing me to the world of research and for supporting me with his advice. Furthermore, I would like to extend my special thanks to Professor Javier Esparza for his invaluable assistance and for taking the time to examine this thesis. My appreciation also goes to Zhuo Cai, for his collaborative efforts on our initial paper and his unwavering support in answering my numerous questions. Lastly, I must acknowledge the endless support and encouragement from my parents, who have been my steadfast supporters throughout my academic journey.

\newpage
\tableofcontents

\newpage
\section{Abstract}

Leader Election (LE) is an important process in distributed systems and blockchain technology, where one of the participants is designated as the leader or coordinator. LE is a part of many distributed systems and blockchain protocols, smart contract applications, and lotteries. Traditional randomized LE methods usually involve distributed random number generation (RNG) to select the leader. These approaches can be vulnerable to manipulation, fail to ensure fairness, or require inefficient and costly procedures, e.g., verifiable delay functions (VDFs) and publicly-verifiable secret sharing (PVSS). 

This Bachelor's thesis in Informatics introduces a new approach to randomized LE. The key insight is that fair leader election can be achieved without relying on explicit decentralized RNG. This is based on the game-theoretic assumption that every participant prefers to be chosen as the leader and would avoid actions that decrease their chances. This game-theoretic incentive can simplify LE protocols compared to RNG. PureLottery, inspired by single-elimination knockout tournaments in sports, offers a fair, bias-resistant, and practical solution for LE in blockchain environments. The underlying principle is that only two participants play against each other in each match, making collaboration efforts to manipulate the result useless.

PureLottery is also efficient with its computational resources and communication complexity, making it practical for implementation as a smart contract. The protocol provides strong game-theoretic guarantees, incentivizing honest behavior, and is robust against adversaries, ensuring that dishonest behavior does not increase their chances of election. PureLottery is designed to be highly resistant to bias, ensuring that each honest player has a minimum winning probability of at least $1/n$, regardless of any potential manipulation by an adversary overseeing the remaining $n-1$ participants. This means that engaging in dishonest practices does not boost an adversary's likelihood of success, even if they have influence over all participants except one. 

Problems similar to the leader election problem can be solved with a modified PureLottery protocol. These problems include a ranking of all participants, electing multiple leaders simultaneously, and leader aversion. The PureLottery protocol has multiple applications in different fields. It can be easily implemented in smart contract applications like lotteries. Moreover, it can be utilized in blockchain and distributed systems protocols. An open-source implementation of PureLottery is provided, dedicated to the public domain.

\section{Publications}
The results of this work were published in the following paper:

\noindent J. Ballweg, Z. Cai and A. K. Goharshady, "\textbf{PureLottery: Fair Leader Election without Decentralized Random Number Generation}," \textit{2023 IEEE International Conference on Blockchain (Blockchain)}, Hainan, China, 2023, doi: 10.1109/Blockchain60715.2023.00051

\section{Introduction}\label{sec:intro}
This work concentrates on random LE within the blockchain framework and introduces a novel leader election algorithm. Leader Election is essential in distributed systems and blockchain technologies because it is used to coordinate tasks among multiple computers, nodes, or users. Leader election protocols have been a central focus in distributed system research for many years~\cite{attiya2004distributed}. LE is about selecting one leader from a group, i.e., these protocols are designed to choose a participant from a set $\{1, 2, \ldots, n\}$ of participants. A sound LE system should be uniform, giving each participant an equal chance to be chosen, and should work without central control.

LE is used in blockchain protocols like in Proof-of-Stake (PoS), where a miner's chance to add a block is proportional to their stake in the currency. It is also used in lotteries and smart contract applications. Traditional LE methods in blockchain often depend on decentralized random number generation (RNG), which can be manipulated, be unfair, or require complex and costly solutions.

This work presents a new viewpoint. We argue that RNG is a more complex problem than LE. Participants naturally want to increase their chances of being chosen in LE, which can be used to create simpler LE protocols than RNG. This is because RNG requires consensus on a random number, which is not necessary for LE.

We propose PureLottery, a new approach to LE. Inspired by knockout tournaments in sports, it does not need decentralized RNG. Instead, it selects a winner through a series of binary competitions. PureLottery is fair, with each participant having an equal chance of winning. It is also resistant to bias, even if almost all other participants are dishonest. Additionally, it is efficient and cost-effective when implemented as a smart contract.

The PureLottery approach is more straightforward than traditional methods, requiring fewer computational resources. We demonstrate that PureLottery is fair, i.e., every participant has a chance of $1/n$ to win, with a uniform random selection of winners, and incentivizes honest behavior. Moreover, it is practical for blockchain environments, needing only a small amount of computational power. For n participants, the number of messages between participants and the smart contract is in $O(1)$ on average and in $O(\log (n))$ in the worst case.

Important terms and foundations are defined in the following. This work uses the terms "player" and "participant" interchangeably.

\subsection{Distributed Systems}
Distributed systems are a type of computing where different computers, often in various locations, work together on a task. It appears to its users as one coherent system~\cite{5832164}. These computers communicate and coordinate their actions by passing messages to one another. Each computer handles a part of the workload, making the process more efficient than using a single computer for the same task. In blockchain systems, however, the aspect of efficiency is often less important. The focus is instead placed on ensuring the absence of a single controlling entity.

\paragraph{Consensus.}
In distributed computing, consensus is the process by which various components of a system reach an agreement on a specific data value or state, crucial for ensuring that multiple computers collaborate effectively and reliably. This agreement is essential in scenarios such as validating transactions within a blockchain or determining the most up-to-date data in a database. Achieving consensus is vital for system integrity, allowing it to function correctly even when some parts fail or encounter errors, and providing resilience against dishonest or malicious attempts to manipulate the system. The challenge of establishing consensus increases with the number of nodes involved, especially in distributed systems where nodes may not inherently trust each other. Various algorithms were developed to achieve this unified agreement, aiming for a consensus mechanism that works efficiently even in complex, multi-node environments. This process of reaching a unanimous decision on a shared state or value among multiple nodes, despite potential failures or security threats, is known as distributed consensus~\cite{Bashir_2020}.

\paragraph{Leader Election.} Leader Election (LE) is a process in distributed computing systems where nodes in a network select a leader among them, i.e., LE is about the selection of a node or a user from a set $\{1, 2, \ldots, n\}$. There can be different criteria of how to select a leader and this depends on the application or the protocol that uses LE. In general (and also in this work) the goal of LE is to select a participant uniformly at random.

\subsection{Blockchain} Blockchain technology operates as a decentralized, peer-to-peer ledger system that is cryptographically protected, designed to be append-only, and immutable, meaning it is exceedingly difficult to alter once data has been recorded. It functions without a central authority, enabling direct interactions and transactions between participants and eliminating the need for intermediaries like banks. Blockchains are implemented on top of peer-to-peer network protocols. The term "append-only" indicates that entries are added to the blockchain sequentially, ensuring that once information is recorded, altering it becomes nearly impossible, effectively rendering the blockchain a secure and tamper-resistant record of transactions. Its immutability can be compromised under specific conditions, such as during a majority attack, but in practice, the blockchain remains immutable and secure~\cite{Bashir_2020} under the assumption of an honest majority and a reliable network.

\paragraph{Distributed Ledger.}
Frequently, the phrases "blockchain" and "distributed ledger" are treated as synonymous. Yet, a notable distinction exists between the two: Blockchains typically organize transactions into grouped blocks as part of their structural design, while distributed ledgers lack this specific arrangement. Aside from this difference, the characteristics of distributed ledgers and blockchains are largely identical~\cite{Bashir_2020}.

\paragraph{Proof-of-Work.} Proof-of-Work (PoW) is a consensus mechanism that requires solving a cryptographic puzzle to add a new block to the blockchain. Bitcoin was the first PoW protocol~\cite{nakamoto2008bitcoin}, introducing a consensus method that does not need a majority of miners to sign a new block and hence allowed further decentralization compared to previous distributed ledgers. PoW makes it very hard to manipulate the state of the blockchain because a dishonest miner needs a large share in the total computation power~\cite{10.1145/3212998}. 

\paragraph{Miner.}
In blockchain technology, a miner uses computing power to solve complex mathematical problems like computing the pre-image of a one-way hash function. This process is part of adding new transactions to the blockchain. By solving these problems, they get the right to append the next block to the chain where they can include validated transactions. Miners are usually rewarded with new units of the blockchain's currency.

\paragraph{Proof-of-Stake.} Proof-of-Stake (PoS) presents a more energy-efficient alternative to Proof of Work protocols. It originated in Peercoin~\cite{Bashir_2020}. In PoS, the privilege to add new blocks to the blockchain depends upon the amount of cryptocurrency a user possesses and is willing to deposit as collateral, emphasizing the concept that participants with a significant stake in the network are less likely to undertake actions detrimental to the system due to their substantial investment. 

\paragraph{Validator.}
Validators are responsible for creating new blocks and validating transactions. Unlike in Proof-of-Work systems, where miners use computational power to mine blocks, in PoS, validators are chosen based on the number of coins they hold and are willing to "stake" as collateral. The more coins staked, the higher the chance of being chosen to validate transactions and create new blocks. 

\paragraph{Node.}
A node refers to any participant connected to the blockchain network. Nodes maintain the network's functionality by keeping, updating, and broadcasting the blockchain's ledger. Not all nodes are miners or validators; some nodes just verify and propagate transactions and blocks without participating in the mining or validating process.

\paragraph{Finality.}
Once a transaction on the blockchain has reached finality, it is regarded as confirmed. A finalized transaction is impossible (or extremely) hard to reverse. A recipient of coins in a transaction will usually wait until it is finalized before they regard the coins to be in their possession.

\paragraph{Smart Contract.}
A smart contract is a program whose execution is guaranteed by the blockchain, ensuring that it operates without interruption and correctly executes the conditions and agreements defined within the contract on the blockchain~\cite{Bashir_2020}. Once deployed on the blockchain, it cannot be taken down. What makes smart contracts useful is that they have a balance that allows them to receive transactions. Coins in the balance can be spent according to the rules defined in the smart contract, which are immutable after its deployment.

\paragraph{Gas Efficiency.}
Each operation in a smart contract requires a certain amount of computational work. On most blockchains (like the Ethereum blockchain), this work is quantified as gas. The resources used for transactions and smart contract execution are measured in gas, and a certain amount of coins must be paid for every gas. Hence, being gas-efficient (in other words, computationally efficient) is a desirable property for smart contracts.

\subsection{Random Number Generation} Random Number Generation (RNG) is the process of creating numbers that cannot be reasonably predicted better than by chance. Random numbers are important for many applications, including cryptography. There are two kinds of randomness: True random and pseudo-random number generators.

\paragraph{True Random Number Generators.} True random number generators rely on physical processes, like electronic noise or radioactive decay, to generate numbers. The numbers are assumed to be truly random because they come from unpredictable natural processes.

\paragraph{Pseudo-Random Number Generators.} Pseudo-random number Generators (PRNGs) use algorithms to produce sequences of numbers that appear random. PRNGs are not truly random because they start from an initial value called a "seed", and if you know the seed and the algorithm, you can predict the numbers. However, to a party that does not possess the seed, the pseudo-random number appears indistinguishable from random.

\paragraph{Uniform Randomness.} in RNG means that each number within a range has an equal probability of being selected. Uniform randomness is important in ensuring fairness and unpredictability in various applications, such as in lotteries and cryptography.

\subsection{Cryptography}
Cryptography uses mathematical data encryption and decryption techniques, ensuring confidentiality, data integrity, authentication, and non-repudiation. Blockchains deemed "cryptographically secure" utilize cryptography to ensure the ledger is safeguarded against alteration and misuse. This involves offering services such as guaranteeing the authenticity of data origin, maintaining data integrity, and preventing denial of transactions~\cite{Bashir_2020}.

Protocols in distributed systems and blockchain systems make extensive use of cryptographic primitives. Cryptographic primitives are basic, low-level algorithms used to construct cryptographic protocols. They are the building blocks of cryptographic systems. Examples that are relevant to this work include one-way hash functions, verifiable delay functions, publicly verifiable secret sharing, and commitment schemes.

\paragraph{One-Way Hash Function.}
Hash functions convert data into a fixed-size hash, which is useful for verifying data integrity. This output is known as the hash value or hash code. A one-way hash function is a hash function that works only in one direction: It is easy to compute a hash value from a pre-image, but it is computationally intractable to generate a pre-image that hashes to a particular value~\cite{Schneier_2015}. This principle is also called non-reversibility. 

\paragraph{Verifiable Delay Function.}
A verifiable delay function (VDF) requires a specified amount of time to compute, but whose results can be quickly and easily verified by others. It is defined as a function \( f: X \rightarrow Y \), which requires the execution of \( f \) through a number of sequential steps, i.e. the process of computing \( f(x) \) cannot be distributed across multiple processors. Assuming that a processor has a maximum frequency for computing operations sequentially, a VDF can guarantee a minimum amount of time to pass until the result is computed. There is an important distinction between the computation and the verification. After the result \( y = f(x) \) has been computed, it can be efficiently confirmed by any observer~\cite{s22197524} without computing the sequential steps again.

\paragraph{Publicly Verifiable Secret Sharing.}
Publicly Verifiable Secret Sharing (PVSS) is a cryptographic method for dividing a secret into different parts (shares) and distributing them among a group of participants. For creating the shares, the threshold $k$ is an important parameter. The threshold $k$ represents the minimum amount of shares that must be combined to reconstruct the original secret. This parameter ensures that the secret can only be reconstructed when a sufficient subset of participants collaborate.

\paragraph{Commitment Scheme.}
A commitment scheme is a secure way to prevent participants from changing a value without revealing the value, i.e., it lets someone commit to a value without revealing it. Later, they can show this value and prove it is the same one they committed to earlier. Often, hashed values are used as a commitment. 

\subsection{Game Theory}
Game theory is a field of study that examines how individuals make decisions in situations where the outcome depends not just on their own actions but also on the actions of others. It is used to analyze strategic interactions where the choices of different players (individuals or groups) influence each other's outcomes. Game theory is often used to analyze protocols in blockchain systems.

\paragraph{Rational Player.} A rational player cares about maximizing their own profit.

\paragraph{Dishonest Player.}
A dishonest participant does not follow the protocol. 

\paragraph{Malicious Player.} A malicious player wants to minimize other player's payoff even if it hurts them. In other words: They might try to minimize other players' payoff even if it is irrational.

\paragraph{Nash Equilibrium.}
A Nash equilibrium in a strategic game is a set of strategies where no single player can benefit by changing their strategy while the other players keep theirs unchanged~\cite{Laraki2019}. In other words, a Nash equilibrium is a situation where, in a game involving two or more players, no player can benefit by changing their strategy while the other players keep their strategies unchanged. This means everyone is doing the best they can, given what others are doing. 

\section{Related Work}
In the next section, we dive into the latest research on various leader election methods, breaking down their complexities and identifying where they might fall short. At the core of many leader election protocols is the use of distributed random number generation (RNG). We'll start by exploring the different methods of RNG, from the naive to more complex strategies designed to patch up their weaknesses. These approaches include commitment schemes, publicly verifiable secret sharing, and verifiable delay functions, which can increase security. Additionally, this section broadens to include online lottery protocols and distributed leader election methods, including the challenge of selecting a single secret leader.

\subsection{Distributed Random Number Generation}
Distributed RNG involves a network of $n$ members who aim to collectively generate a uniformly random number and agree on the result. Since participants lack trust, the protocol must be designed to prevent any individual from manipulating or skewing the outcome. While a robust distributed RNG system inherently offers a solution for distributed Leader Election, the reverse may not always be true.

\paragraph{Naive Solution.}
A basic approach to Random Number Generation (RNG) is for each participant \( i \in \{1, ..., n\} \) to select a random value, such as a random string \( x_i \) from the set \( \{0,1\}^l \). These values are then shared among the group, and the collective output \( y \) is calculated as \( y = x_1 \oplus x_2 \oplus \cdots \oplus x_n \), where \( \oplus \) represents the bitwise exclusive or (XOR) operation. If every value is independently chosen and at least one value is truly from a uniform distribution, then the final output \( y \) will be a uniformly random string. 

However, in practical scenarios, this method is significantly flawed due to the non-simultaneous nature of message transmission. The participant who announces their value last, after the other \( n-1 \) members, can manipulate the final output. If participant \( n \) is the last to submit their value and intends for the output to be \( d \) from the set \( \{0, 1\}^l \), they can set \( x_n \) as \( x_1 \oplus x_2 \oplus \cdots \oplus x_{n-1} \oplus d \). This choice guarantees that the final outcome \( y \) will be \( d \).

\paragraph{Commitment Scheme.} 
A commitment scheme is often employed to counteract the last participant's dominance and mimic simultaneous actions. This cryptographic primitive enables a sender to commit to a specific value, maintaining its secrecy from the receiver. Conversely, the receiver is equipped to validate the committed value later, provided the sender reveals it. The commitment scheme operates in two stages. During the \emph{commit} phase, the sender possesses a message \( x \) and selects a random string \( r \) from \( \{0,1\}^\kappa \). The sender then encodes these into \( c \) and transmits \( c \) to the receiver. Typically, a hash function is applied here, resulting in \( c:= \texttt{hash}(x, r) \). The sender transmits a hint string \( k \) to the receiver in the subsequent reveal phase. With \( k \) in hand, the receiver can unlock the commitment \( c \) to extract and confirm \( x \). For instance, in the context of a hash-based commitment, we could use the hint $k:= (x, r)$. Formally, a commitment scheme is expected to fulfill the following two security properties:

\begin{itemize}
    \item \emph{Hiding:} Upon receipt of a commitment \( c \), it should not reveal any details about the message \( x \). To put this formally, for every \( x_0 \), \( x_1 \), define \( p_0 \) as a distribution \( \{(r,c) | r \overset{{\scriptscriptstyle\$}}\gets \{0,1\}^\kappa, c \overset{{\scriptscriptstyle\$}}\gets \mathbf{Commit}(x_0, r) \} \), and \( p_1 \) as \( \{(r,c) | r \overset{{\scriptscriptstyle\$}}\gets \{0,1\}^\kappa, c \overset{{\scriptscriptstyle\$}}\gets \mathbf{Commit}(x_1, r) \} \). The distributions \( p_0 \) and \( p_1 \) should then be computationally indistinguishable.
    \item \emph{Binding:} Different values than \( x \) must not result in the identical commitment \( c \). In more technical terms, for every probabilistic polynomial time algorithm that is not uniform and generates \( x_0 \), \( x_1 \) along with \( r_0 \), \( r_1 \), the chance that \( x_0 \neq x_1 \) but \( \mathbf{Commit}(x_0, r_0) = \mathbf{Commit}(x_1, r_1) \) should be a negligible function relative to \( \kappa \), which is the length of \( r_0 \) and \( r_1 \). 
\end{itemize}

Using a commitment scheme, RNG participants can secure their values in the commit phase and disclose them in the reveal phase once all have committed. During the commit phase, the final participant is unaware of others' values and thus cannot manipulate the output deliberately. Nevertheless, a dishonest participant might opt not to disclose their value in the reveal phase, and it is practically impossible to unveil the commitment independently. If the protocol finalizes the output without their value, this effectively skews the distribution. For a random bit, by committing to \(1\), a participant can control the output by choosing whether to reveal. 

For example, consider a commitment scheme for random number generation for three participants that is used to elect a leader. Participant 1 is honest while Participants 2 and 3 are colluding and dishonest. Every participant $i$ has to commit to a number $x_i \in {0, 1, 2}$ and reveal it later. The resulting random number $x = \sum_{i=0}^2 x_i \mod 3$ is computed by adding all revealed numbers. If the two colluding participants chose their values $x_1 = 1$ and $x_2 = 2$, they can always get $x$ to be their desired outcome, no matter what value player 1 chooses. Depending on player 1's revealed value, either just player 2, just player 3, or both have to reveal their committed values. In this example with three players, the two dishonest players always win. In other examples, the advantage gained from cheating may be less drastic but persisting to encourage cheating.

Implementing a penalty system, like forfeiting a pre-agreed deposit for such acts, might not deter this if the economic gains from influencing the output surpass the loss of deposits. For instance, making a lottery that uses a commit-reveal scheme game theoretically secure would require participants to pay a deposit as high as the lottery reward, which is impractical and unrealistic in most applications. Additionally, restarting the protocol when a participant withholds revelation can also lead to biased outcomes. The last participant might calculate the value and withhold revelation unless the result benefits them. 

\paragraph{Publicly-Verifiable Secret Sharing.}
A different approach to address the problem of harmful tampering involves the use of a publicly-verifiable secret sharing method, as discussed in~\cite{berry:simple-pvss,bryan:scalable-bias-resistant-distributed-randomness}. In this method, during the initial phase, each participant selects a secret value and distributes parts of this secret to the others. Subsequently, in the second phase, the participant has the option to disclose their secret, allowing others to verify its authenticity. Alternatively, if a participant decides to withhold their secret, the other participants can join forces to reconstruct it using the shared parts. However, the drawback of the PVSS approach is its quadratic communication complexity. Moreover, it presupposes that most participants are trustworthy, a necessary condition for the accurate reconstruction of the secrets when faced with dishonest entities.

\paragraph{Verifiable Delay Functions.} play a crucial role in enhancing the tamper-resistance of RANDAO, a collection of Ethereum smart contracts designed to generate random numbers in a distributed manner~\cite{randao}. VDFs, as outlined in~\cite{boneh:vdf}, are functions that require a predetermined length of sequential, non-parallelizable processing to produce a result. This design ensures that the outcome can only be determined after a specific time, unaffected by the use of powerful parallel processors. Once computed, VDFs allow for the creation of an efficient proof of the result, which eliminates the need for others to repeat the resource-intensive computation to verify the output. 

In the context of RANDAO, applying a VDF to the XOR result of a random number generator prevents dishonest participants from biasing the output, as they cannot foresee the VDF's result during the reveal phase. However, this approach introduces challenges: It leads to gas-inefficient computations on the blockchain for result verification and necessitates substantial off-chain computation for running the sequential algorithm to evaluate the VDF. 

Moreover, it is important to note that the output of a VDF is not guaranteed to be uniformly distributed. This uniformity assumption holds only if we consider hash functions as ideal random oracles. This introduces a broader issue with VDFs, hash functions, and pseudo-random number generators: their inability to guarantee completely random and uniform outputs.

\subsection{Online Lottery Protocols}
A different group of research focuses on online lottery systems that do not necessarily rely on blockchain technology. For instance, Chow's protocol~\cite{DBLP:conf/iccsa/ChowHYC05} features an online lottery scheme that utilizes a hash chain to connect the lottery tickets of participants. In this scheme, a Verifiable Random Function (VRF) is used to generate verifiable randomness from the hash chain. Additionally, a Verifiable Delay Function (VDF) is employed to prevent last-minute manipulation by dishonest participants. However, similar to commitment schemes incorporating VDFs, this approach raises concerns regarding efficiency due to the time required for sequential processing in VDFs. Moreover, it does not ensure uniform randomness in the lottery outcomes.

Lee's design~\cite{DBLP:journals/csi/LeeCI09} incorporates the Chinese Remainder Theorem and blind signatures~\cite{DBLP:conf/crypto/Chaum82} in lottery tickets. To pick a winner, Lee's system uses a pseudo-random number generator, seeding it with the cumulative modular sum of encrypted random values provided by all participants. However, this method is vulnerable to collusion between the lottery dealer and the last participant, who could potentially manipulate the random seed by decrypting the values of other participants. 

Liu improved on this by changing the random seed to a result based on Lagrange interpolation, derived from all participants' random values, which are initially committed~\cite{DBLP:journals/ijcomsys/LiuLCCJ14}. Despite this enhancement, Liu's scheme still needs to solve the problem of dishonestly concealing random values, a common issue in commitment schemes.

Grumbach's approach~\cite{DBLP:conf/dais/GrumbachR17} utilizes delay functions to combat manipulation. This method stands out for its use of a Merkle tree structure, enabling quick probabilistic verification in large-scale lotteries. This tree is different from our contribution.

Lastly, Xia proposed a lottery scheme using symmetric bivariate polynomials for sharing random secrets among various lottery centers~\cite{DBLP:journals/symmetry/XiaLHC19}. This distributed randomness approach is akin to distributed RNG methods seen in publicly-verifiable secret sharing schemes.

The methods described previously are centralized in nature, involving a central authority or dealer who collects values from all participants. However, in numerous instances, this dealer role could be substituted with a smart contract, resulting in approaches that align more closely with the decentralized methods discussed in the subsequent section.

\subsection{Blockchain-Based Lottery Schemes}
Blockchain technology and smart contracts are highly effective for creating decentralized protocols that do not rely on trusted intermediaries. As a result, numerous lottery and Leader Election schemes have been developed using smart contracts. According to current literature~\cite{DBLP:conf/icbc2/CaiG23,DBLP:journals/corr/abs-1912-00642,DBLP:conf/blockchain2/PanZLWS22,DBLP:conf/icbc2/ChenHCW19,DBLP:conf/desec/LiZ019, DBLP:journals/corr/abs-1912-00642}, all existing blockchain-based lottery systems employ random number generation. This RNG is either generated through pseudo-random number generators, through an external decentralized RNG protocol, or through random oracles that utilize blockchain states as seeds. Consequently, these systems share the inherent limitations of the RNG methods they employ. It is important to note that elements of the blockchain, like block hashes, are at risk of being manipulated by dishonest miners who might aim to influence the LE outcome. Moreover, using RNG with VDFs or hashes derived from blockchain states is not guaranteed to be uniformly random and, hence, can come with uniformity issues. While some of these systems offer additional features like enhanced privacy, they cannot overcome the fundamental limitations of RNG protocols. 

Moreover, works have been published for specific use cases of leader election, such as for selecting validators in proof-of-stake protocols~\cite{zhuo-marble, DBLP:conf/icbc2/CaiG23}. These approaches deliver game-theoretic incentives to follow the protocol honestly. 

The work most similar to this work was published by Miller and Bentov~\cite{7966964}. It introduces a leader election algorithm without utilizing distributed random number generation and aligns with the basic version of the PureLottery algorithm, i.e., working for $2^{i}$ participants. However, it does not include a solution for how to deal with an arbitrary number of participants in a fair manner.

\subsection{Single Secret Leader Election}
Research in another area focuses on the challenge of selecting a single secret leader~\cite{boneh2020single}. This process involves two key aspects: (i) only one leader is chosen from the group of participants, and (ii) the identity of the leader remains unknown to others until the leader decides to disclose their success. An example of this can be seen in~\cite{backes2022framework}, which employs a tree-based, multi-round protocol akin to our tournament approach. It covers scenarios where participants have varying probabilities of winning (weighted setting),~\cite{backes2022framework} uses a tree representation for each participant's input, with a depth of $O(\log (s))$. Here, it is assumed that the weights are integers and $s$ is the total weight. In comparison, PureLottery adapts well to the weighted setting without additional complexities. Our protocol, however, does not prioritize keeping the leader's selection secret, resulting in significantly lower computational and communication complexities.

\section{The PureLottery Protocol}
This section introduces the novel PureLottery protocol. We denote the number of participants in the lottery with $n = 2^m$. We will extend the protocol to an arbitrary number of players in Section~\ref{sec:arbitrary-many-players}.

PureLottery operates without the need for decentralized Random Number Generation. There is not one random implying the leader but a random result in every match that a participant goes through. Hence, our protocol avoids the need for consensus-based random numbers that blend inputs from various parties. 

The LE protocol operates via a smart contract and is fully decentralized, meaning all participants have equal roles and capabilities. It is presumed that each participant is linked to an account on the blockchain network, possessing both a secret key and a public key, and capable of executing standard operations like hashing and encryption.

\paragraph{Deposit.}
A deposit by the players is optional because the protocol is already well-incentivized, as no advantage can be gained by not following the protocol. However, signing up for the LE process should be restricted by some criteria to avoid a Sybil attack. Thus, it should not be possible to obtain arbitrarily many tickets to increase the probability of winning. For instance, in a lottery, access would be restricted through a ticket price. As another example, in a leader election process used in a proof of stake protocol, registration would be restricted by the amount of stake.
The PureLottery protocol only works if there is an incentive to be the winner. Otherwise, there is also no incentive to follow the protocol, i.e., using PureLottery to determine a participant who gets punished would not align with the incentives used in the protocol.

\paragraph{Lottery Context.} Our implementation focuses on the lottery context. Nevertheless, the protocol can be applied to different applications. In the context of a lottery implemented through this LE protocol, the selected individual becomes the winner and is entitled to the entire prize amount. All contributions designated for the lottery are awarded to the elected leader, who simultaneously becomes the lottery winner. Other participants, who did not win, cannot claim any portion of the money. 

\paragraph{Intuition.}
This work is fundamentally guided by the insight that Leader Election is often an easier challenge than Random Number Generation, owing to inherent motivations for participants to boost their selection likelihood. For instance, in RNG, a participant might exit the protocol early, impacting the randomness of the outcome. In contrast, in LE, participants face a clear choice: either finish the protocol or forfeit the chance of being elected. This dynamic allows for simpler LE solution designs, leveraging these incentives, as compared to complex RNG methods. 

The reason for the desirable security properties of the PureLottery protocol is that only two players compete against each other in every match. Hence, there is no third player who can collaborate with another player and influence the game by not revealing. Only having matches between two players leaves no advantage in not revealing.

\subsection{Two-Player Case}

In our PureLottery protocol, the scenario with two players aligns with traditional methods that utilize commitment schemes. However, when expanding to a larger number of players, our approach differs significantly. For ease of explanation, we initially discuss the basic scenario with only two participants, that is, $m=1$. In this setting, one of these two participants needs to be uniformly elected as the leader. Following the typical structure of commitment schemes, the PureLottery protocol is divided into two stages: A commitment phase and a revelation phase. Within our smart contracts, time is gauged by block numbers, with each phase corresponding to a specific block interval.

\paragraph{Commit Phase.}

During the commitment phase of the PureLottery protocol, participants must register and transmit a commitment message to the contract. The commitment from participant $i$ takes the form $\langle \texttt{hash}(x_i || r_i) \rangle$, where $\texttt{hash}$ represents a pre-defined cryptographic hash function, $x_i \in \{0,1\}$ is a randomly chosen bit, and $r_i \in \{0,1\}^\kappa$ is a random string acting as a salt to conceal $x_i$. The symbol $||$ signifies the concatenation of strings. Participants have the option to register and commit simultaneously or through two separate messages to the contract. The process can also be divided into two sub-stages, initially allowing all participants to register and then requiring them to commit after the registration deadline.

Participants are limited to sending a single commit message in this phase. Sending multiple messages or failing to send a valid one results in being labeled as dishonest. Such participants are disqualified from the LE, defaulting the leadership to the other player. Participants are numbered based on the sequence of their registration with the contract, a sequence that is clear-cut and derived from the transaction order on the blockchain. In cases where the protocol operates without a blockchain, such as for Proof of Stake, participants can be organized based on their public key or its hash. This ordering does not confer any advantage to the participants.

\paragraph{Reveal Phase.}
Similar to a standard commitment scheme, both participants in the PureLottery protocol are required to send a reveal message to the contract during the reveal phase. The reveal message from participant $i$ should consist of $\langle x_i, r_i \rangle$. The contract then calculates $\texttt{hash}(x_i || r_i)$ and checks its consistency with the initial commitment of participant $i$. If the resulting hash values do not align, the reveal message is disregarded, and the revelation is deemed invalid. A participant who fails to reveal correctly and within the allocated time frame (during the reveal phase) automatically relinquishes the leadership to the other participant.

Once the reveal phase concludes, if a participant has not committed or revealed in time, their opponent is declared the winner. If neither participant acts dishonestly, the contract determines $y = x_1 \oplus x_2$. The victory goes to participant $i$ if $y \equiv i - 1 \mod 2$. Specifically, the first participant wins if $y = 0$, while the second participant wins if $y = 1$.

It is important to note that in this basic protocol, an honest participant who genuinely selects $x_i$ from a uniform distribution always has at least a 50\% chance of winning, regardless of the opponent's actions. A dishonest opponent can only increase the winning probability of the honest player by opting out of the game. Therefore, no participant who is rational in terms of game theory would stray from the Nash equilibrium strategy of choosing an $x_i$ from the uniform distribution.

\begin{figure}[H]
\centerline{\includegraphics[width=0.4\columnwidth]{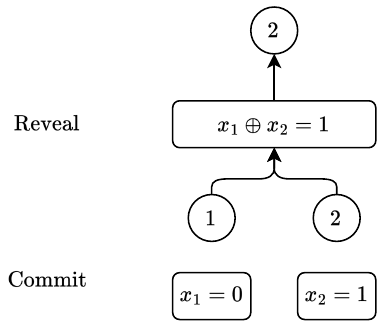}}
\caption{An example of a two-player execution of PureLottery.}
\label{fig: tournament_2_players_example}
\end{figure}

\begin{figure}[H]
\centerline{\includegraphics[width=0.4\columnwidth]{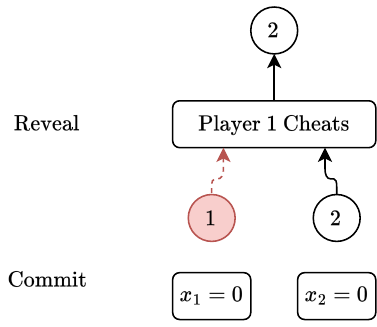}}
\caption{An example of a two-player execution of PureLottery with an honest and dishonest player.}
\label{fig: tournament_2_players_cheat_example}
\end{figure}

\paragraph{Examples.}
Examples are shown in Figures \ref{fig: tournament_2_players_example} and \ref{fig: tournament_2_players_cheat_example}.

\subsection{The Case with $n=2^m$ Players}
We now introduce the PureLottery protocol for a scenario involving $n = 2^m$ players. This protocol draws a parallel with a knockout-style tournament, common in sports like football. It builds upon the fundamental concept of a two-player model and extends over $m$ rounds. Throughout these rounds, pairs of neighboring players engage in a commit-reveal game, as outlined in the prior section. In this setup, the defeated player is removed, and the winner advances. In cases where neither player succeeds, meaning both breach the protocol, a placeholder or dummy player is introduced who then moves forward but is always designed to lose in subsequent rounds. With the player count reducing by half after each round, the process culminates in the selection of a single leader by the conclusion of the $m$ rounds.

\paragraph{Registration and Commitment.} 
In the initial stage of the extended PureLottery protocol, the process is similar to the commit phase described earlier. Every participant, labeled as $i$, selects $m$ random bits, denoted as $x_i^1, x_i^2, \ldots, x_i^{m}$, that are uniformly distributed. Additionally, they pick $m$ random salts, represented as $r_i^1, r_i^2, \ldots, r_i^{m}$. Subsequently, they calculate the corresponding hashes:

$$
\begin{array}{c}
h_i^{m} = \texttt{hash}(x_i^{m} || r_i^{m})\\
h_i^{m-1} = \texttt{hash}(x_i^{m-1} || r_i^{m-1} || h_i^{m})\\
\vdots\\
h_i^{j} = \texttt{hash}(x_i^{j} || r_i^{j} || h_i^{j+1})\\
\vdots\\
h_i^{1} = \texttt{hash}(x_i^{1} || r_i^{1} || h_i^{2}).
\end{array}
$$

Our goal is to utilize the hash $h_i^j$ in the $j$-th round of the protocol. The final hash, $h_i^{m}$, acts as a commitment to the random bit $x_i^{m}$, similar to the method described previously. However, each preceding hash $h_i^j$ commits not only to $x_i^j$ but also to the subsequent hash $h_i^{j+1}$. Therefore, any change of the $x_i^j$ values would result in a different $h_i^1$. Essentially, $h_i^1$ represents a commitment to all bits chosen by player $i$, offering the benefit of needing to submit only one hash value instead of hashes for each round. 

In the commit phase, participants must send a message to the smart contract, including a deposit and the commitment $h_i^1$. Similar to the earlier process, a player who fails to submit the deposit or a valid commitment is deemed to have lost and is replaced with a dummy player for subsequent rounds. Players are sequenced based on their registration time with the contract. This ordering is arbitrary and can be substituted with a sequence based on public keys or any other definitive method. Altering the order does not provide any player with an advantage.

\paragraph{Reveal Rounds.}
After the commit phase, half of the players get eliminated in every round during the reveal phase. The elected leader is the sole player left after the round $m$. The reveal phase starts, spanning $m$ rounds, each with a set deadline. Before round $j$, there are $n_{j-1} = \frac{n}{2^{j-1}}$ players left, and after round $j$, there are $n_j = \frac{n}{2^{j}}$ players left. These participants, in a sorted sequence, are paired up for competition. Each player $i$ must send a message to the smart contract containing $\langle x_i^j, r_i^j, h_i^{j+1} \rangle$. The contract verifies these inputs against the previously declared $h_i^j$ for player $i$, checking if it matches $\texttt{hash}(x_i^{j} || r_i^{j} || h_i^{j+1})$. A mismatch leads to the dismissal of the reveal message.

If player $i$ is paired with player $i'$ (where $ i'> i$), failure by either player to correctly reveal results in their losing the round, and the other player wins. If both reveal correctly, the contract calculates $y_{i, i'} = x_i^j \oplus x_{i'}^j$. Victory goes to player $i$ if $y_{i, i'} = 0$ and to player $i'$ if $y_{i, i'} = 1$.

\begin{figure}[H]
\centerline{\includegraphics[width=0.8\columnwidth]{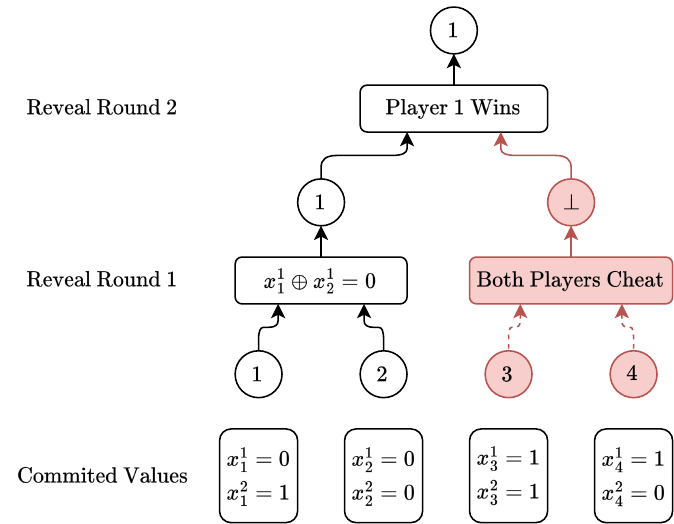}}
\caption{An example of a PureLottery execution with four players where two players cheat and a dummy player $\perp$ is introduced.}
\label{fig: tournament_4_players_both_cheat_example}
\end{figure}

In cases where both players $i$ and $i'$ fail to reveal, they both lose. Here, a dummy player with the same identifier $i$ is introduced for the next round. These dummy players, managed by the smart contract, automatically lose against any real player who reveals correctly but win in cases where their opponent does not reveal. This mechanism ensures a consistent halving of participants in each round, maintaining $n_j$ players after round $j$. Figure \ref{fig: tournament_4_players_both_cheat_example} shows this mechanism.

\begin{figure}[H]
\centerline{\includegraphics[width=0.8\columnwidth]{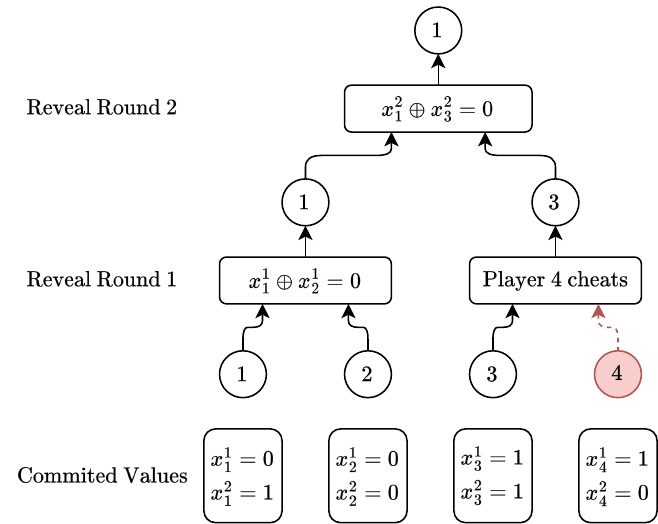}}
\caption{An example of a PureLottery execution with four players.}
\label{fig: tournament_4_players_example}
\end{figure}

\paragraph{Examples.} 
Figure \ref{fig: tournament_4_players_example} illustrates a PureLottery example with four participants. At the start, the players pick their random bits $x_i^j$ and salts $r_i^j$ and commit to these values. In the initial round, player $1$ competes against player $2$, and player $3$ faces player $4$. Player $1$ wins against player $2$ as both have $x_1^1$ and $x_2^1$ equal to $0$, resulting in $y_{1, 2} = 0$. In the other match, player $3$ discloses $x_3^1 = 1$. Knowing his impending loss due to $x_4^1$, player $4$ chooses not to reveal, leading to his default loss for non-revelation. Players $1$ and $3$ advance to the subsequent round, where the smart contract already holds $h_1^2$ and $h_3^2$ from the prior round. Now, each player from $\{1, 3\}$ must reveal $x_i^2$ and $r_i^2$. The contract verifies these hashes and calculates $x_1^2 \oplus x_3^2$, identifying player $1$ as the final winner and elected leader.

Figure \ref{fig: tournament_4_players_both_cheat_example} presents a different scenario where both players $3$ and $4$ do not reveal their values in the first round, resulting in their elimination. A dummy player, represented by $\perp$, moves to the next round. In the match between player $1$ and $\perp$, player $1$ wins if it correctly reveals its value. In essence, $\perp$ is predetermined to lose unless its real opponent cheats.

\subsection{Three-Player Case}
In the previous section, for simplicity and conciseness, we described the PureLottery protocol assuming the number of players $n$ is a power of 2 and defined $m := \log_2 n$. This assumption simplifies the presentation and analysis of the protocol's security and complexity. However, this condition can be relaxed without impacting the protocol's correctness or the validity of our analysis. In order to generalize the algorithm for an arbitrary number of players, we start with the case of $n=3$ players.

\paragraph{Naive Approaches.}
If three players participate in the leader election, a naive approach would be to make a match with three participants. However, if the basic unit includes three participants, there is a chance that player 1 and player 2 might conspire. In the reveal phase, these colluding participants could strategize based on the actions of the honest player 3. Their options include: (i) revealing both $x_1$ and $x_2$, (ii) revealing $x_1$ but concealing $x_2$, or (iii) revealing $x_2$ but concealing $x_1$. If any participant fails to reveal their value without causing the protocol to halt, then one of the other two will win. This way, the colluding party may increase their winning probability through deceitful collaboration. Essentially, the protocol would suffer from the same vulnerabilities as a simple commitment scheme for random number generation. It becomes apparent that in every match, only two players should play against each other. Solving this problem is the characteristic that sets PureLottery apart, offering benefits that are unattainable with traditional random number generators based on commitment schemes, even though they may seem similar in basic instances.

\begin{figure}[H]
\centerline{\includegraphics[width=0.4\columnwidth]{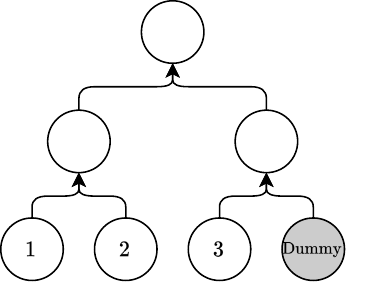}}
\caption{The non-perfect binary tree becomes a perfect binary tree by adding a dummy player.}
\label{fig: add_dummy_player}
\end{figure}

Creating a tournament structure for three players where only two players compete in a match would result in an unfair advantage for one player. If winning probabilities are the same in every match, the only fair tournament tree would be a perfect binary tree. We could use dummy nodes to fill up the tree so that it is a perfect binary tree again (as shown in Figure \ref{fig: add_dummy_player}). However, the dummy player does not solve the problem of unfairness no matter which of the following two approaches are taken. One way could be to let real players automatically win against dummy players, but the players who compete against a dummy player get an unfair advantage. Another possibility is to assign a random number to the dummy players based on the values that all other players submitted. However, this approach comes with the same challenges as any distributed random number generation protocol. It is prone to players that influence the value by not revealing it. Hence, we set aside the approach of using dummy players.

\begin{figure}[H]
\centerline{\includegraphics[width=0.7\columnwidth]{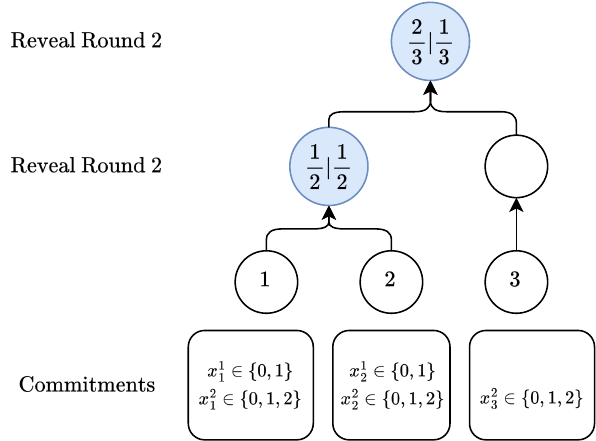}}
\caption{An example of the PureLottery setup with three players. The fractions in the blue bubbles denote the winning probabilities for the left and right children, respectively.}
\label{fig: tournament_3_players_setup}
\end{figure}

\paragraph{Skewed Commitment Game for Three Players.}
Because dummy players do not help, we have to deal with a leader election tree that is not a perfect binary tree, i.e., not every node has two children or the leaves are on different levels (as shown in the example of Figure \ref{fig: tournament_3_players_setup}). 
We can achieve fairness in such a tree by adjusting the winning probabilities in every node. The game should be set up so that in the final match, the player $1$ and player $2$ have a $2/3$ chance of winning, while player $3$ has only a $1/3$ chance of winning. This adjustment ensures that each of the players $1$, $2$, and $3$ have an equal probability of reaching the final round from the starting point. The final match works differently compared to previous examples. For this match, every player has to choose and commit a value $x_i^2 \in \{0, 1, 2\}$. The match result will be computed using modular addition, instead of XOR. If the result is 0 or 1, the left player wins. If the result is 2, player $3$ wins. We call this a "skewed commitment game" and generalize this approach in the next section to work with an arbitrary number of players.

\subsection{The General Case}\label{sec:arbitrary-many-players}
\begin{figure}[H]
\centerline{\includegraphics[width=0.9\columnwidth]{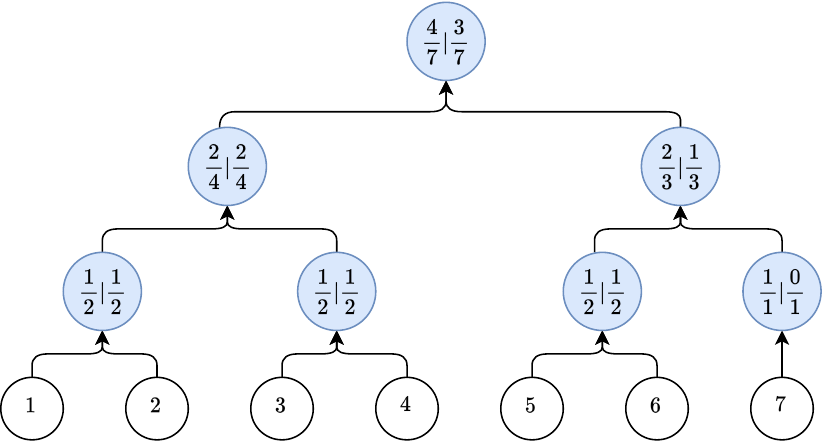}}
\caption{An example of a PureLottery setup for seven players.}
\label{fig: tournament_7_players_setup}
\end{figure}

To make PureLottery for an arbitrary number of players, we have to adjust the winning probabilities in every match depending on the leaves below the respective node. This balancing can be achieved easily. For any internal node $u$ in the tree, denote $k_u$ as the number of leaves that are descendants of $u$. Let $l_u$ and $r_u$ represent the number of descending leaves from $u$'s left and right child, respectively, so that $k_u = l_u + r_u$. We assign the probabilities $\frac{l_u}{k_u}, \frac{r_u}{k_u}$ to node $u$, indicating the chances that the winner at $u$ comes from the left or right subtree. Refer to Figure~\ref{fig: tournament_7_players_setup} for illustration. The task is to design a game between the two players that realizes these specific winning probabilities. This is referred to as a skewed commitment game.

\paragraph{Skewed Commitment Game.} Instead of every tree node containing a $1/2$ winning probability for both players, the winning probability is proportional to the number of leaves below each node if every player has the same chance of winning. We call this a skewed commitment game. Imagine a game where player $1$ is meant to win with a probability of $l/k$, and player $2$ with a probability of $r/k$, where $r = k-l$. Initially, each player $i$ selects a random integer $x_i$ from $\{0, 1, \ldots, k-1\}$ uniformly and commits to this number. In the next phase, they disclose their committed numbers, and the sum $y_{1,2} = x_1 + x_2 \mod k$ is calculated. Player $1$ wins if $y_{1,2} < l$, otherwise player $2$ wins.

This skewed commitment game can be integrated into our PureLottery protocol. The primary difference to the approach for $2^m$ players is the use of modular addition instead of XOR, and the $x_i^j$ values are no longer mere bits but larger numbers. Specifically, if the ancestor of player $i$ at round $j$ has $k$ descendant leaves, then $x_i^j$ should be within $\{0, 1, \ldots, k-1\}$. For instance, in Figure \ref{fig: tournament_7_players_setup}, player $3$ should choose $x_3^1$ from $\{0, 1\}$, $x_3^2$ from $\{0, 1, 2, 3\}$, and $x_3^3$ from $\{0, 1, \ldots, 6\}$.

\paragraph{Commitments.}
Instead of committing to all random values before the first round, the user only commits to the random value for the first round. With every reveal message, the user then also sends a new commit value for the subsequent round. It is important to build check mechanisms into the protocol so that a user can only reveal the current round's value and commit to the next round's value.

\begin{figure}[H]
\centerline{\includegraphics[width=1\columnwidth]{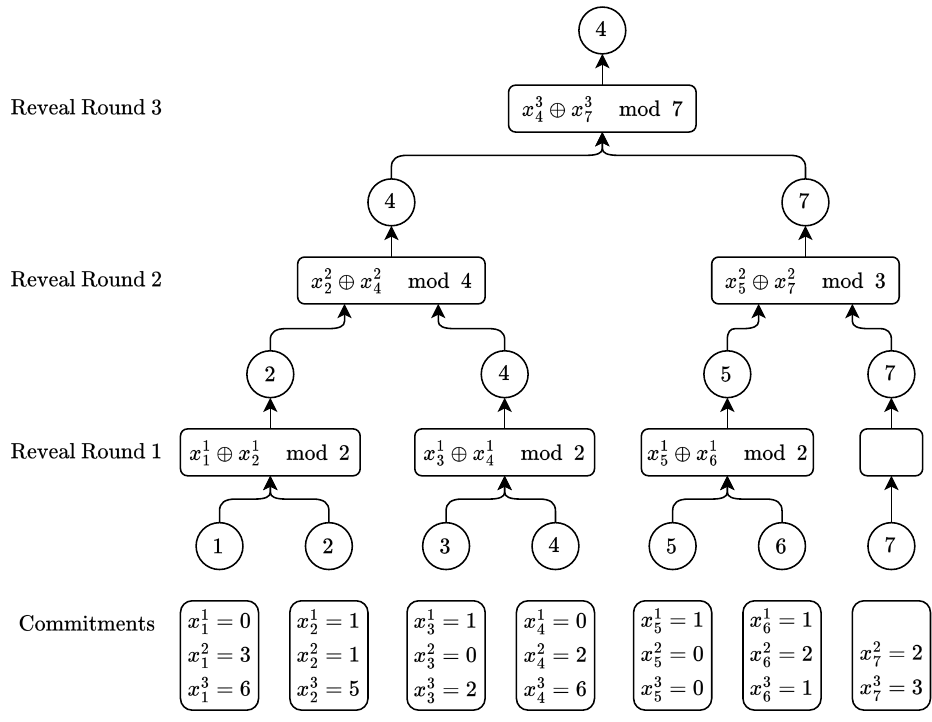}}
\caption{An example of a PureLottery execution for seven players.}
\label{fig: tournament_7_players_example}
\end{figure}

\paragraph{Example.}
Using the tournament setup from Figure \ref{fig: tournament_7_players_setup} and playing through it results in the results of Figure \ref{fig: tournament_7_players_example}.

\subsection{Extension to Arbitrary Winning Probabilities}\label{sec:pure-lottery-arbitrary-probabilities}
The mechanism described in the previous section can be used not only for an arbitrary number of participants but also for assigning arbitrary winning probabilities. Suppose we have a lottery where every participant can buy an arbitrary number of tickets. A participant buying more tickets should also have a higher probability of winning. This is the setting used in proof-of-stake protocols because every player has a winning probability proportional to their stake. 

We assign every participant an integer $w_i$ and call it weight. The higher the weight, the higher the winning probability. In this setting, the value $k_u$ of any node u is the sum of all weights belonging to the leaves descending from $u$. In the same way, $l_u$ and $r_u$ are the sum of weights from $u$'s left and right descending leaves. This way, every participant can get an arbitrary rational number $\frac{w}{\sum_{i=1}^{n} w_i} \in \mathbb{Q} $ assigned as their probability of becoming the leader. For every node $u_j$ above the player's leaf until the root, the player computes $k_{u_j}$ and commits to a number $x_i^j \in \{0, ..., k_{u_j}\}$. In the revealing phases $y_{i,i'} = x_i^j + x_{i'}^j \mod k_u$ is calculated. Assuming that $i < i'$, player $i$ wins if $y_{i,i'} < l$, otherwise player $i'$ wins.

\begin{figure}[H]
\centerline{\includegraphics[width=0.8\columnwidth]{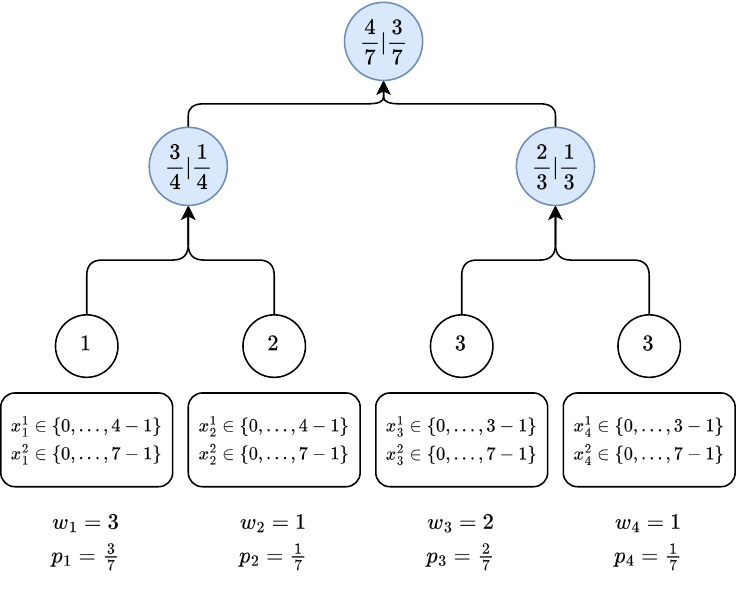}}
\caption{An example of a PureLottery setup with seven players, where every player has a different probability of winning. The weights $w_i$ could stand for the number of lottery tickets a player has bought. $p_i$ denotes the probability of a player becoming the leader.}
\label{fig: tournament_4_players_arbitrary_probability_setup}
\end{figure}

\begin{figure}[H]
\centerline{\includegraphics[width=0.7\columnwidth]{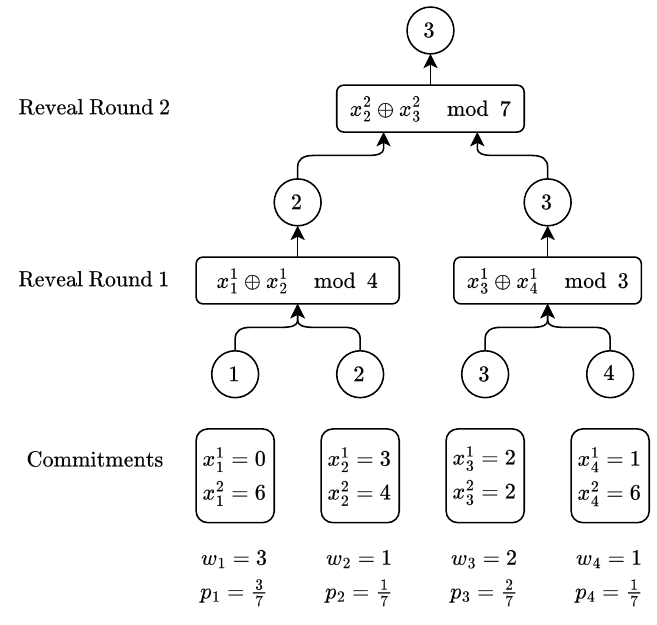}}
\caption{An example of a PureLottery execution with four players and arbitrary winning probabilities. It is based on the setup from Figure \ref{fig: tournament_4_players_arbitrary_probability_setup}.}
\label{fig: tournament_4_players_arbitrary_probability_example}
\end{figure}

\paragraph{Examples.}
Examples of a PureLottery setup and execution with arbitrary winning probabilities are shown in Figures \ref{fig: tournament_4_players_arbitrary_probability_setup} and \ref{fig: tournament_4_players_arbitrary_probability_example}, respectively.

\section{Implementation}
We implemented the PureLottery protocol for an arbitrary number of players and arbitrary winning probabilities as an Ethereum smart contract, written in the Solidity language, can be accessed at 
\[\href{https://doi.org/10.5281/zenodo.10116210}{https://doi.org/10.5281/zenodo.10116210}.\]
It is implemented as a lottery and uses the most advanced version of the PureLottery algorithm with an arbitrary number of users and arbitrary winning probabilities, as introduced in Section~\ref{sec:pure-lottery-arbitrary-probabilities}.

The contract runs in three phases: Signup, commit-reveal, and end. During signup, players pay for a lottery ticket, the tournament tree is created, and leaves are added with every additional player that signs up. The players also have the option to resign, which means they leave the lottery and get a refund. For determining winners of matches later, the valid range of random values for every match has to be known to the smart contract. To keep the implementation simple, we implement a tree that keeps track of the weights in every node. 

In the commit-reveal phase, every player commits to the random value for the next round and reveals the random value of the current round. During every reveal-commit phase, there are breaks to ensure the blockchain has reached finality so that nothing can be changed or added to the previous submission phase. The committing and revealing actions are implemented in the same function in the smart contract to signal that both actions have to be executed at the same time. During the initial commitment, the function ignores the revealed values and during the last reveal, the function ignores the next commitment. In the end phase, the elected leader can withdraw the contract balance (the sum of all ticket fees). 

In the following, all public function signatures that a user can execute by interacting with the smart contract are listed. For more details please refer to the source code.

\begin{center}
\begin{minipage}{0.89\textwidth}
\begin{lstlisting}
function signup(uint numberOfTickets) public payable 

function resign() public payable signUpStage returns(uint numberOfTickets)

function commitReveal(uint randomNumber, uint salt, bytes32 commitmentNextRound) public

function payout() public payable
\end{lstlisting}
\end{minipage}
\end{center}

In the remaining parts of the section, we perform a theoretical runtime complexity analysis, followed by an experimental gas cost analysis.

\subsection{Runtime Complexity}

\paragraph{Communication Costs.} 
In distributed systems, it is desirable to have as little communication as possible to make the system reliable, e.g. because the message delivery fails. In the PureLottery protocol, each participant sends a maximum of $m+2 = O(\log n)$ messages to the contract, which is the upper limit in case a participant is the winner. This total comprises one registration message, a commitment message containing $h_i^1$, and a maximum of $m$ messages across various reveal phases. The elected leader, who remains in the game until the end, must send a message in every round. 

On the other hand, many participants are eliminated in the initial rounds, thus reducing the number of messages they need to send. For any given participant, the expected number of rounds they stay in is less than $2$, calculated as $1 + 1/2 + 1/2^2 + ... + 1/2^m$. Consequently, on average, each participant sends fewer than $4$ messages of constant length to the contract. 

\paragraph{Gas Costs.}
Gas costs in blockchains usually consist of the computation costs and storage costs for the variables that are stored in the smart contract. The gas fees for executing a function are minimal, typically $O(1)$. Besides the signup function, each function in the smart contract requires $O(1)$ computation on the chain. 

\begin{figure}[H]
\centerline{\includegraphics[width=0.7\columnwidth]{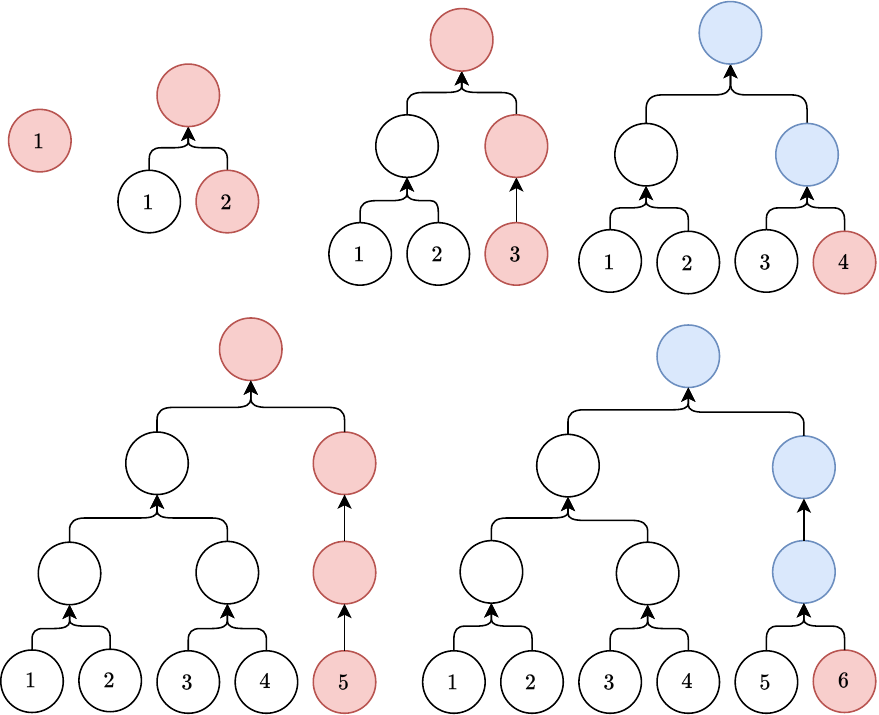}}
\caption{Signup process and creation of the tournament tree. A red node stands for a node that is added by the new player. A blue node has to be updated by the new player.}
\label{fig: signup_tree_creation}
\end{figure}

The signup function requires $O(\log (i))$ operations for every player $i$, where $i-1$ players have already signed up. This is illustrated in Figure~\ref{fig: signup_tree_creation}.  The total gas expenditure for the signup function in our implementation is thus $\Theta(\sum_{i=1}^{n} \log(i)) = \Theta(\log(n!))$. It is theoretically possible to implement the signup function with $O(1)$ operations. However, this would increase the complexity of the smart and introduce cases where a user has to pay high gas fees.

The overall theoretical gas expenditure for all participants is $\Theta(n)$, which is efficiently minimal since registering every participant alone consumes this amount of gas. The contract is also designed for gas efficiency, mainly performing hash operations, which are low-cost on programmable blockchains like Ethereum~\cite{wood2014ethereum}. High gas-consuming operations like VDFs, PVSS, and zk-SNARKs have been deliberately avoided in the contract design.

\paragraph{Off-chain Computation.} 
In the PureLottery protocol, participants are required to execute $O(\log n)$ hash operations to calculate the $h_i^j$ values. These computations can be carried out off-chain on the participant's device.

\subsection{Experimental Gas Cost Analysis}
We split the experimental gas cost analysis into two parts: The signup process and the execution of the actual PureLottery algorithm. During the signup, the contract creates a tournament tree, as previously illustrated in Figure~\ref{fig: signup_tree_creation}. Creating a variable that is stored in the contract is an expensive operation that costs a lot of gas. Updating a variable, however, is cheaper. This is why the players who have to create new (red) nodes have to pay more gas than the players who just have to update existing (blue) nodes during the signup process. The following chart illustrates this with the peaks directly after a power of two at signup positions 3, 5, 9, 17, and 33.

\begin{figure}[H]
\centering
\begin{tikzpicture}
\begin{axis}[
    title={Gas for Signup by Player},
    xlabel={Signup Order},
    ylabel={Gas Used},
    ymajorgrids=true,
    grid style=dashed,
    scaled y ticks=false,
    yticklabel style={/pgf/number format/fixed},
    ymin=0,
    ymax=270000,
    width=0.9\textwidth,
    height=0.6\textwidth,
]
\addplot table [
    col sep=comma, 
    x=player, 
    y=gas_for_execution] 
    {data/signup_gas_costs.csv};

\newcommand{\baselinedepth}{20000}
\draw [gray, dashed] (axis cs:2,\baselinedepth) -- (axis cs:2,\pgfkeysvalueof{/pgfplots/ymax}) node[pos=-0.09, anchor=south] {2};
\draw [gray, dashed] (axis cs:4,\baselinedepth) -- (axis cs:4,\pgfkeysvalueof{/pgfplots/ymax}) node[pos=-0.09, anchor=south] {4};
\draw [gray, dashed] (axis cs:8,\baselinedepth) -- (axis cs:8,\pgfkeysvalueof{/pgfplots/ymax}) node[pos=-0.09, anchor=south] {8};
\draw [gray, dashed] (axis cs:16,\baselinedepth) -- (axis cs:16,\pgfkeysvalueof{/pgfplots/ymax}) node[pos=-0.09, anchor=south] {16};
\draw [gray, dashed] (axis cs:32,\baselinedepth) -- (axis cs:32,\pgfkeysvalueof{/pgfplots/ymax}) node[pos=-0.09, anchor=south] {32};

\end{axis}
\end{tikzpicture}
\label{fig:signup_gas_costs}
\end{figure}

Next, we show the gas costs of the core part in the PureLottery protocol - the commit-reveal phase and the final payout for the winner. The exact results can differ slightly depending on the exact order in which participants reveal. It is important to note, that the signup phase was excluded from the following measurements.

The winner of the leader election goes through all tournament phases and hence has to pay the most gas. Gas costs for the winner depending on the number of participants are shown in the following chart.

\begin{figure}[H]
\centering
\begin{tikzpicture}
\begin{axis}[
    title={Gas for PureLottery Execution of Winning Player},
    xlabel={Number of PureLottery Participants},
    ylabel={Gas Used},
    ymajorgrids=true,
    grid style=dashed,
    scaled y ticks=false,
    yticklabel style={/pgf/number format/fixed},
    ymin=0,
    ymax=550000,
    width=0.9\textwidth,
    height=0.6\textwidth,
]
\addplot table [
    col sep=comma, 
    x=number_of_players, 
    y=winner_total_gas] 
    {data/winner_total_gas.csv};

\end{axis}
\end{tikzpicture}
\label{fig:winner_total_gas}
\end{figure}

As described in the theoretical results, the total amount of gas used by the PureLottery algorithm is linear in the number of participants. 

\begin{figure}[H]
\centering
\begin{tikzpicture}
\begin{axis}[
    title={Gas for PureLottery Execution for all Players},
    xlabel={Number of PureLottery Participants},
    ylabel={Gas Used},
    ymajorgrids=true,
    grid style=dashed,
    scaled y ticks=false,
    yticklabel style={/pgf/number format/fixed},
    ymin=0,
    ymax=6000000,
    width=0.9\textwidth,
    height=0.6\textwidth,
]
\addplot table [
    col sep=comma, 
    x=number_of_players, 
    y=sum] 
    {data/tournament_total_gas_costs.csv};

\end{axis}
\end{tikzpicture}
\label{fig:tournament_total_gas_costs}
\end{figure}

\section{Security Analysis}

\subsection{Fairness and Uniform Randomness.}
In the following discussion, we first consider a scenario with $n=2^m$ participants in the leader election protocol and then extend the discussion to an arbitrary n, where only one individual will be ultimately chosen as the leader. Provided that every participant complies with the protocol, each participant possesses an equal chance of selection as the winner, denoted by the probability $1/n$.

It is important to highlight that, unlike our method, pseudo-random number generators lack a formal assurance of producing outputs with a perfectly uniform distribution. Consequently, the methodologies previously employed fall short of offering an absolute uniform random distribution in determining the winner, a limitation our approach overcomes.

If all participants are honest, they all send valid messages in each round. The system always gets correct reveal messages and uses the XOR operation to determine a single winner and loser in each pair. In these scenarios, since each participant picks a bit \(x_i^j\) randomly, the XOR outcome is a random bit as well. The chance for a participant to make it through each reveal round is \(1/2\). With different random values for each round, the probability of surviving all \(m\) rounds is \(1/2^m = 1/n\), which is the fair probability we want. It is important to note that the fairness of each round does not rely on both participants being honest. If one participant is honest and reveals their bit correctly, they have a minimum chance of \(1/2\) to move to the next round. Additionally, when a portion of participants fails to transmit, commit, or reveal messages accurately, their chances of success are strictly decreased. 

The same logic applies when \(n\) is not a power of 2. In this general case, every participant picks a random value from ${0,...,k_u}$. The addition $\mod k_u$ is uniformly at random if at least one of the values in a match is chosen uniformly at random. Here, a player's chance of winning is the product of probabilities from the leaf to the root, and all these products equal \(1/n\), as designed.

It is in the participant's best interest to choose the random values uniformly to make it as unpredictable as possible. The Nash equilibrium in the XOR game is selecting values from a uniform distribution.

\subsection{Strong Bias-Resistance.}
Regardless of potential dishonest collaboration, a group of $t \leq n-1$ participants is unable to enhance their collective chances of winning the prize. To elaborate, should $t$ participants comply with the rules, their likelihood of producing a winner within their ranks stands at $t/n$. Our framework ensures that no coalition of $t$ participants can engage in deceit to boost their cumulative odds of victory to $t/n+\varepsilon$ for any $\varepsilon>0$. Furthermore, we affirm that each honest participant maintains a minimum chance of $1/n$ to become the winner, despite facing $n-1$ scheming opponents—that is, in a scenario where all other contenders are engaged in dishonest conduct aimed at disadvantaging the honest contestant.

To show strong bias-resistance, let us first examine the scenario when \( n=2 \). In a situation where both participants engage in dishonesty, it can be demonstrated that their combined probability of winning does not surpass \( 1 \) because none of them becomes a winner. Conversely, if both are honest, it has been previously established that their winning probabilities are equal, each at \( 1/2 \). Now, imagine a case where player 1 maintains honesty while player 2 acts as a dishonest opponent.

\begin{itemize}
    \item The opponent may choose their value $x_2$ in a non-uniform way. Nevertheless, because commitment schemes are secretive, the opponent lacks knowledge about $x_1$, which is selected randomly. This means that $x_2$ is picked without any influence from $x_1$. As a result, $y = x_1 \oplus x_2$ will have a uniform distribution regardless of how $x_2$ is distributed.
    \item In the scenario where the adversary fails to disclose truthfully, the adversary loses the game, and the winning probability becomes $0$. Additionally, the adversary relinquishes their deposit. Under these circumstances, given that player $1$ adheres to the protocol honestly, his chances of winning the lottery exceed $1/2$, specifically becoming $1$. 
\end{itemize}
The same analysis is equally applicable to a skewed commitment game, confirming that an adversary is unable to reduce the winning probabilities of an honest player.

We demonstrate that the PureLottery protocol, with $n=2^m$ participants, maintains strong bias-resistance, i.e. resistance to bias even if $n-1$ adversaries collaborate. This is shown in the following by ensuring each honest participant has a minimum chance of $1/2$ to advance in each reveal round. In situations where two adversaries collaborate, they can decide the round's winner, likely choosing to optimize their combined chances of selection. Nonetheless, if one of these adversaries later faces an honest player $i$ in round $j$, they cannot predict player $ i$'s value $x_i^j$, which remains concealed until revealed by player $i$ in round $j$. Therefore, if player $i$ is honest and selects $x_i^j$ randomly, its chances of winning any round remain unaffected. 

This principle also applies when $n$ is not a power of two. In every round, a participant has an assigned minimum chance of advancing to the next round. If two adversaries collaborate and decide who the winner is, they cannot improve their combined chances of winning because they cannot predict the values of an honest player that they might face later. Hence, the honest player's chances of winning remain unaffected.

To summarize, having several separate reveal rounds in the design guarantees that for honest participants, each round's competition between two players is unaffected by earlier rounds. The adversary's value is fixed before the current round, but the honest participant's value remains undisclosed until the round begins. As a result, every honest participant has at least a $1/n$ chance of being chosen. In essence, any dishonest behavior only lowers the adversaries' winning chances. Consequently, no rational player or group of players would choose to act dishonestly, assuring that every player is elected with an exact probability of $1/n$.

\subsection{Liveness.}
Liveness means that no group of $t<n$ participants can halt the protocol before its due completion. The concept of liveness is intricately linked to the resistance against bias. As an illustration, in the context of random number generators that employ commitment schemes~\cite{randao}, it is possible for opponents to skew the results by exiting the protocol prematurely during the reveal phase. Alternatively, some random number generators adopt publicly verifiable secret sharing (PVSS) to prevent the protocol from being discontinued mid-way~\cite{bryan:scalable-bias-resistant-distributed-randomness}. Yet, the effectiveness of PVSS hinges on the honesty of the majority of participants. As such, the assurance of liveness in these protocols is effectively safeguarded only when facing a fraction of $t<n/2$ adversaries.

The PureLottery protocol, however, continues even if a participant acts dishonestly, either by sending incorrect messages or not revealing values. Such participants are immediately out of the game. This approach is secure due to the protocol's strong bias resistance. 

A dishonest individual or group might attempt to skew the leader selection's probability distribution, yet this would not work to their advantage. Therefore, in practical terms, no rational participant or group would withhold their number revelation. This creates a solid game-theoretical foundation, as revealing is always the most advantageous strategy. By not following the protocol, the most an adversary can achieve is to pass their winning chances to their immediate opponent in the pairing. When making decisions, they lack information about future rounds beyond what they control themselves. Thus, even in collusion with their opponent, such dishonesty is unproductive and actively discouraged due to the loss of deposit.

With a little change in the protocol, it is possible to change from a game-theoretic liveness guarantee to a liveness guarantee even if $n-1$ players collaborate maliciously, i.e. they are not acting rationally and do not care about losing money. The situation where our described protocol with the dummy players might not be sufficient looks like this: In one round, no player reveals. This is not in their interest, but it is theoretically possible and would keep the protocol from the goal of electing a single leader. We can solve this by letting the players who lost in the previous round move up and continue the leader election process with them.

\section{Related Problems}
In this section, we introduce some problems that are similar to LE.
Hence, we can modify the PureLottery approach to solve these related problems while preserving the same security properties.

\subsection{Ranking}
A trivial solution for a ranking of players would be to execute the leader election $n-1$ times in a row with all remaining players. However, this would consume much time and also increase the gas costs on the blockchain. It is possible to create a ranking of all participating players in $m$ rounds using only one tournament. 

In a ranking, we create a sequence $(p_{j_1}, ..., p_{j_n})$ from a set of participants $\{p_1, ..., p_n\}$. We use the PureLottery version for $2^m$ players as it is the most straightforward. It can be generalized to an arbitrary number of players. The basic idea here is that a ranking can be created if all players who lose at the same stage will play in a smaller leader election against each other. 
How it works:
\begin{itemize}
  \item All players have a fixed starting order.
  \item The rules are the same as in the PureLottery algorithm: Two players commit and reveal a random number to determine the winner. A player who does not reveal will lose the current round's match.
  \item The winners proceed into the next round. The losers from one round start another tournament. Their ranking position will be better than the previous losers but worse than the winners from the current round.
\end{itemize}

\begin{figure}[H]
\centerline{\includegraphics[width=0.7\columnwidth]{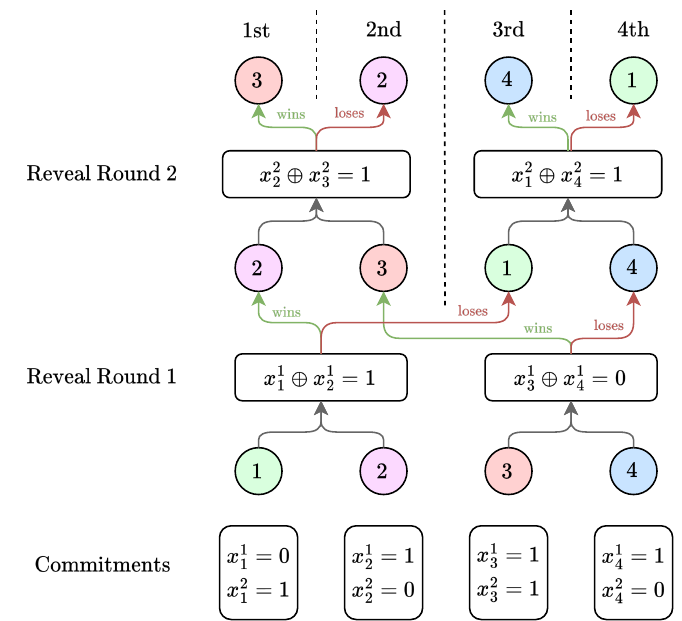}}
\caption{An example of the PureLottery ranking execution with four players.}
\label{fig: ranking_4_players_example}
\end{figure}

\paragraph{Example.} An example tournament is shown in Fig. \ref{fig: ranking_4_players_example}. 

\paragraph{Complexity.}
To rank all players with the described protocol, every participant has to go through a complete tournament, similar to an LE winner going from the leaf to the root in the PureLottery protocol. Thus, the communication complexity is $O(\log(n)$ for every participant who honestly follows the protocol. Consequently, the total communication costs are in $O(n\cdot \log(n))$ messages. 

\subsection{Multiple Leaders}
Selecting $z$ leaders is a problem related to PureLottery. Different approaches to selecting multiple leaders are introduced in this section.

\begin{figure}[H]
\centerline{\includegraphics[width=0.7\columnwidth]{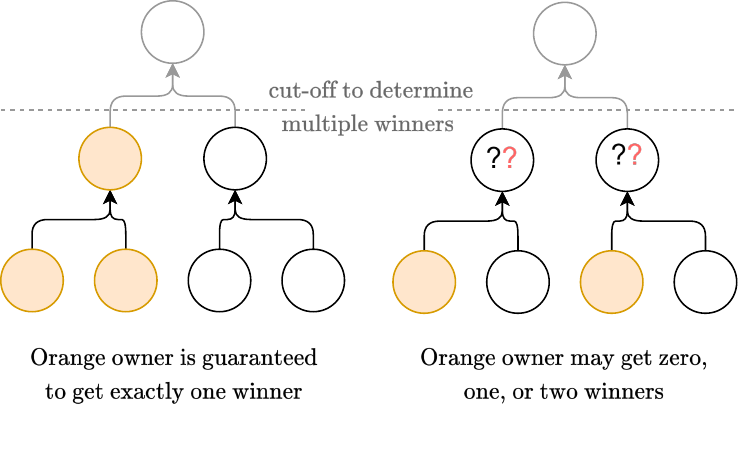}}
\caption{An example that demonstrates how an individual exerting control over multiple participants can significantly impact who the winners will be.}
\label{fig: multiple_winners_vulnerability}
\end{figure}

\paragraph{Naive Version.}
Selecting multiple leaders can be done by terminating the leader election process early. For instance, if $z=2$ leaders should be selected, the PureLottery protocol is stopped one reveal phase earlier. If $z=4$ leaders should be selected, the PureLottery protocol must be stopped two reveal-phases before the end. The leader election process is fair in the sense that the average number of wins cannot be influenced. However, this simple approach is affected by the winning outcome variability problem, which is relevant in some applications. The winning outcome variability problem means that if a participant controls multiple players, they can influence the winning outcome variability. This means that depending on the order of signing up, the participant who controls various players can change the outcome. The setup is the following (see Figure \ref{fig: multiple_winners_vulnerability}): A tournament has four players, and one person controls two participants. Two participants are supposed to become leaders. In this case, the person controlling two participants can influence the game through the signup order. If the two controlled participants sign up as the first two (or similarily the last two) participants, one of them will for sure become a leader. Other signup orders leave this open, and the result might be that zero, one, or both participants controlled by the same instance become leaders.

\paragraph{Multiple Tournaments.} It is possible to circumvent this problem by having $z$ tournaments in a row. In every tournament, only one leader is elected. The next tournament then consists only of participants who are not leaders yet. However, this increases the total execution time by a factor $z$ compared to the normal PureLottery protocol. Moreover, the communication complexity and gas cost are also increased by a factor of $z$.

\begin{figure}[H]
\centerline{\includegraphics[width=0.9\columnwidth]{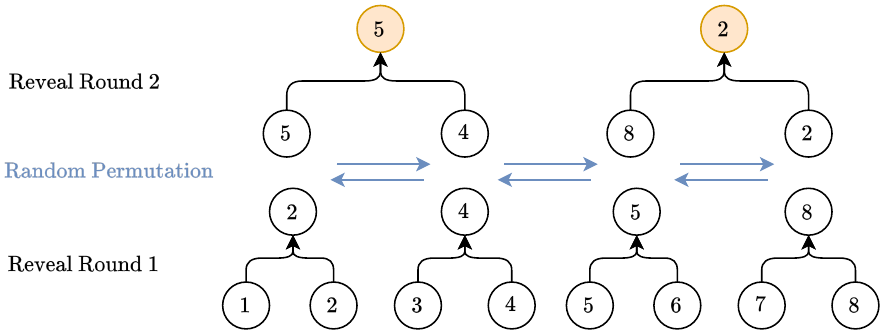}}
\caption{An example of determining multiple winners and solving the outcome variability problem with random permutations applied between the revealing phases. Commitments and revelations are left out for simplicity and readability.}
\label{fig: multiple_winners_permutation}
\end{figure}

\paragraph{Random Permutations.}
Another approach is to apply random permutations in every round to solve the variability problem. After one tournament round, the player positions are reshuffled based on the revealed random values from the previous round. More precisely, when a round ends, all revealed random values from that round are taken to create a uniform random number $r$. The resulting random number is used to create a permutation, e.g. with a shuffle algorithm or by applying a uniform hash function $r$ times to the player positions. This approach only works if the lottery has at least two rounds of revealing because the permutation can only take place between two consecutive rounds. See Fig. \ref{fig: multiple_winners_permutation} for an example.

\paragraph{Parallelization.} 
To improve the duration of selecting the leaders, it is possible to have multiple leader election tournaments in parallel where each tournament elects one leader. This works similarly to the PureLottery protocol with the difference that every message that a participant sends to the contract contains the commit and reveal data for the same round in all the $z$ tournaments. A participant might be elected in multiple tournaments. To avoid repetitive elections of the same participant, the affected participant gets de-elected in all other tournaments. More precisely, if a participant is the elected leader in one of the $j \in {1, ...,z}$ tournaments, they get de-elected in all subsequent tournaments $j, ..., z$. In these subsequent tournaments $j, ..., z$, the player becomes the new leader who played the most recent match against the de-elected leader. 

\begin{figure}[H]
\centerline{\includegraphics[width=0.95\columnwidth]{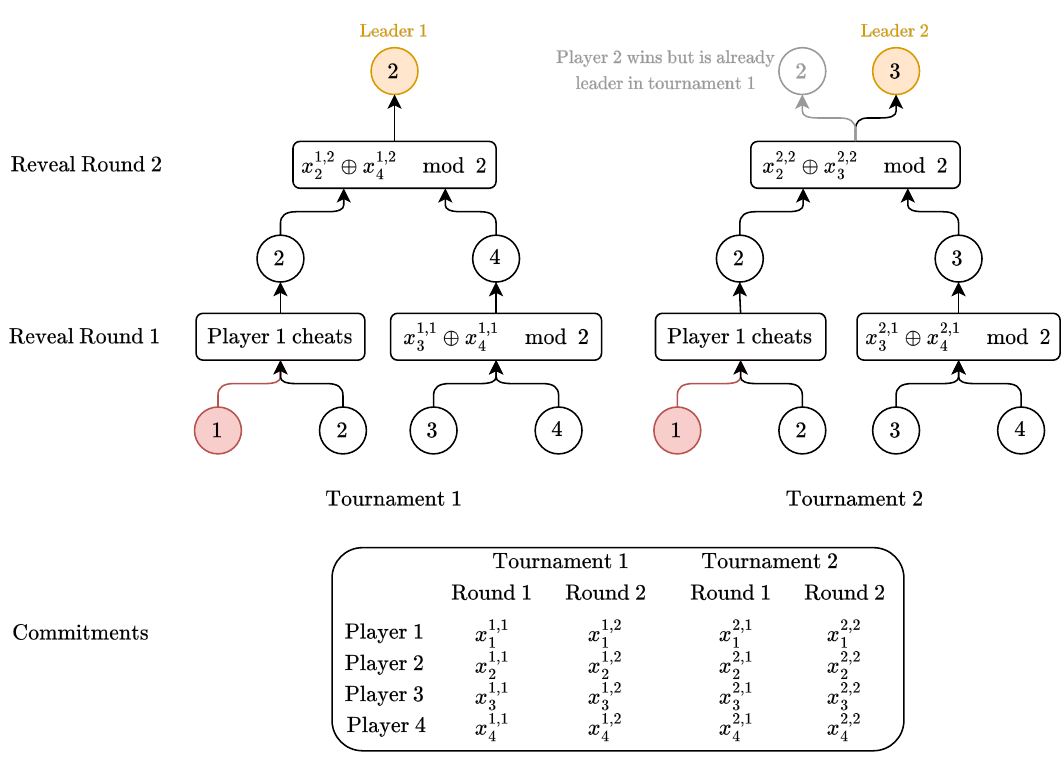}}
\caption{An example of executing the parallelized PureLottery protocol with two tournaments. Tournaments 1 and 2 are denoted in the first digit in the superscript.}
\label{fig: multiple_winners_parallalization}
\end{figure}

To prevent skewing the tournaments through withholding, the smart contract receives one message for all parallel tournaments. This means that if the player decides to cheat in one tournament, they get disqualified in all tournaments. For instance, if a player in one round reveals correctly in Tournament 1 but not correctly in Tournament 2, they get disqualified in both tournaments. See Figure \ref{fig: multiple_winners_parallalization} for an example. Compared to PureLottery, the gas and communication costs are $z$ times as high compared to the PureLottery protocol. However, compared to executing multiple tournaments in a row, the time needed is the same as in the PureLottery protocol.

\subsection{Leader Aversion}
Leader aversion, a converse to normal LE, involves a negative incentive to become the leader. More precisely, in some applications, we may have to perform a leader election with the difference that there comes a disadvantage through being the elected loser. We call this "leader aversion" or "negative leadership contest". For instance, imagine a situation where a group of friends are in a restaurant. Because no one desires to pay the bill, the idea of selecting a person randomly is proposed, utilizing a mechanism similar to the PureLottery protocol.
The goal is to leave the tournament as fast as possible. The rules here change a bit:
\begin{itemize}
    \item The winner of every match leaves the tournament, and the loser stays in.
    \item If a participant does not reveal their committed value, they stay in the tournament.
    \item If two players in a match do not reveal, one of them is chosen to go into the next round, based on previously defined criteria (e.g., first/last signup order or higher/lower public key).
\end{itemize}

\begin{figure}[H]
\centerline{\includegraphics[width=0.7\columnwidth]{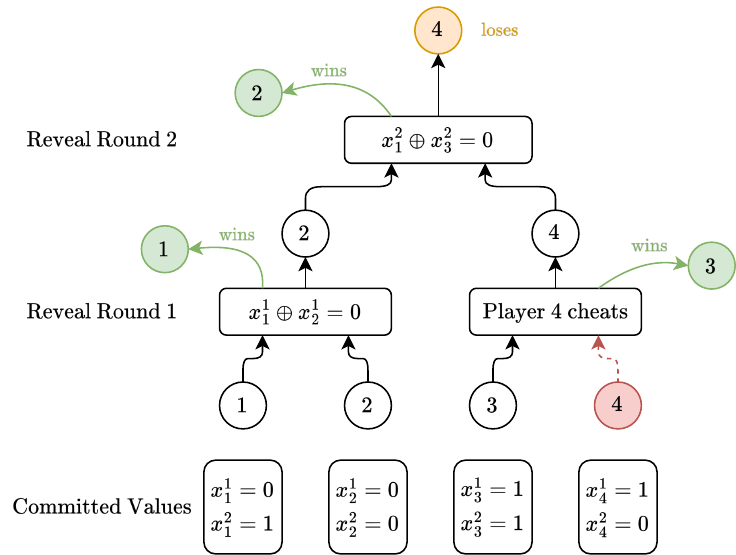}}
\caption{This example shows how four players go through a negative leadership contest.}
\label{fig: leader_aversion_4_players_example}
\end{figure}

\begin{figure}[H]
\centerline{\includegraphics[width=0.7\columnwidth]{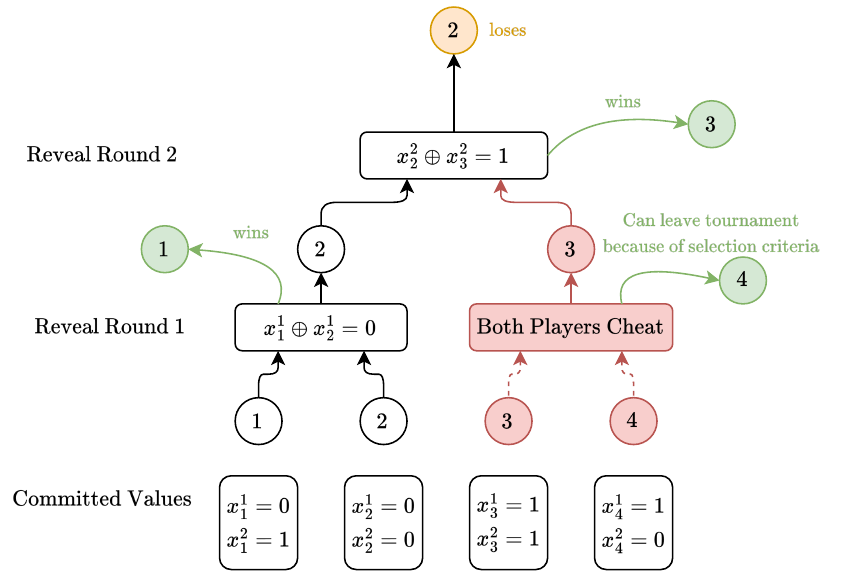}}
\caption{Another example of four players going through a negative leadership contest. The selection criteria for two players who do not reveal is that the player with a smaller number is chosen to continue in the tournament. Player 3 does not reveal in round 1. Later in round 2 player 3 decides to reveal (e.g. because it came back online), and wins the last round.}
\label{fig: leader_aversion_4_players_2_cheat_example}
\end{figure}

\paragraph{Examples.}
Two examples are shown in Figures \ref{fig: leader_aversion_4_players_example} and \ref{fig: leader_aversion_4_players_2_cheat_example}.

\paragraph{Alternative.} There is a different version of a leader aversion protocol, which is more simple. It is useful when it is sufficient to elect not exactly one but at least one negative leader.
\begin{itemize}
    \item If all participants reveal honestly, exactly one participant is chosen.
    \item Otherwise, all cheaters are elected and have to pay.
\end{itemize}

\section{Real-World Applications}\label{sec:applications}
The PureLottery leader election protocol is ideally suited for a variety of applications that require fair and randomized leader election, especially in scenarios where the participants mutually distrust each other. The protocol's transparency allows for external verification of the election process. If all messages between participants are accessible, or if the procedure is implemented using a smart contract with data stored on the blockchain, the fairness of the leader election can be independently verified.

\subsection{Smart Contract Applications}
Our PureLottery protocol is easily adaptable to smart contract implementations, opening up a multitude of applications.

\paragraph{Lotteries.}
Besides gambling, lotteries have many practical applications. Essentially, a lottery means that several players participate in the lottery while one or multiple players will get the prize through a fair and randomized election. This prize can be anything. For example, a useful real-world example would be a lottery for student dorms. In many cities, the demand for student dorm rooms far exceeds the supply. Hence, many administrations decide to assign dorm rooms to students by drawing the lots.

As a consequence, there is often unfairness involved in the process because students with personal connections to the administration might be preferred. In extreme cases, someone might even try to bribe the administration. A distributed leader election protocol like PureLottery could make sure that everyone is treated fairly and would make the paying of bribes useless.

School placement processes can also benefit from the PureLottery protocol, ensuring equitable allocation of school slots in over-subscribed institutions.

Another application of lotteries is online gaming, where PureLottery can be used to randomly select players for making critical game decisions, ensuring an unbiased selection process. PureLottery could bring fairness and transparency to processes that are currently opaque.

\paragraph{Financial and Election Audits.}
Our protocol offers a novel approach to enhancing fairness and transparency in crucial auditing processes, such as financial and election audits.

In financial audits, it is noteworthy that certain tax authorities incorporate random selection into their auditing procedures~\cite{Hashimzade2016Predictive}. This method is particularly used to determine which taxpayers are subject to audit, promoting an unbiased approach to compliance checks.

Regarding election audits, the protocol's application is particularly significant in post-election scenarios. Risk-limiting audits, for instance, benefit from the random selection of ballot samples~\cite{Ottoboni2018Risk-Limiting}. This randomness is essential for maintaining the statistical integrity of the audit, ensuring that the audited sample accurately represents the entire voting population. Furthermore, some regions adopt random selection in forming independent electoral audit committees. This strategy is aimed at minimizing potential biases or political influences in the audit process, thereby upholding the audit's objectivity and credibility~\cite{Carson2009Appointments}.

\paragraph{Public Funding.}
In public funding initiatives, PureLottery could play a crucial role in the random selection of projects. Recent trends among science funding policy experts suggest a growing preference for integrating lotteries with traditional peer review in grant evaluation. This shift is driven by the belief that incorporating randomness can more effectively identify and support innovative research ideas~\cite{Shaw_2023}.

\paragraph{Decentralized Autonomous Organizations.}
Decentralized Autonomous Organizations, often referred to as DAOs, function primarily or in part through the management of smart contracts, as highlighted by Wang et al. in 2019~\cite{wang2019decentralized}. These types of organizations empower their members with the ability to vote, an activity that takes place on the blockchain. In DAOs, each member typically has equal voting rights. However, to propose new initiatives or start the voting process, a coordinator must be elected from among the members. Therefore, practically all DAOs in actual operation necessitate the use of LE protocols.
Similarly, in collaborative projects, the protocol can help in deciding the leader for group decisions.

\subsection{Blockchain Protocols}
\paragraph{Proof of Stake Consensus.}
Protocols based on proof-of-stake, referenced in studies like~\cite{algorand,ouro,praos}, blend various methods to select new miners. PureLottery offers a solution to the drawbacks found in various proof-of-stake consensus mechanisms. Unlike methods dependent on blockchain elements or computationally intensive processes like VDFs or PVSS, PureLottery ensures a fair and random leader election without susceptibility to manipulation or excessive computational demands.

\paragraph{Byzantine Fault Tolerance.}
In the realm of Byzantine Fault Tolerance (BFT) protocols, PureLottery emerges as a key tool for ensuring unbiased and random leader selection. This aspect is crucial in maintaining the integrity of systems, especially in scenarios where dishonest nodes might be present. Consider the PBFT model, where leader selection is conducted in a sequential, Round-Robin manner~\cite{Zou2020Bycon}. Here, participants are expected to be honest, utilizing timeouts to supervise the current leader and transition their support to the next leader upon timeout occurrence~\cite{Abraham2016Solidus}. While this approach is effective, the introduction of PureLottery's randomness can significantly bolster the system's resilience against potential stability threats.

\paragraph{Federated Byzantine Agreement Protocols.}
BFT systems typically operate within a permissioned network where all nodes are recognized and predetermined for participating in the consensus process. In contrast, Federated Byzantine Agreement (FBA) offers a more flexible and scalable approach to trust. Popularized by Stellar's blockchain network~\cite{10.1145/3341301.3359636}, FBFT thrives on a federated model, freeing nodes from the need to trust all other network participants uniformly.

In the context of FBA protocols, as exemplified by Stellar, the application of PureLottery offers a fair and impartial method for selecting nodes in charge of validating transactions~\cite{Losa2019Stellar}. This integration not only enhances the fairness in node selection but also aligns with the inherent principles of decentralized and unbiased operation, which is crucial for blockchain networks.

\paragraph{Sharding.}
The PureLottery protocol finds significant application in sharding mechanisms within blockchain networks. Sharding is a process that divides a blockchain network into smaller, more manageable pieces, or "shards", each capable of processing its own set of transactions and maintaining a portion of the overall state. This division necessitates a fair and transparent method for electing leaders within each shard, who are responsible for validating and processing transactions~\cite{9284731}. The PureLottery protocol ensures that this leader selection is conducted in a manner that is random, unbiased, and verifiable, thereby enhancing the efficiency and integrity of sharding in sharded blockchain systems.

\paragraph{Rollups.}
Similarly, in the context of Layer 2 scaling solutions such as Plasma~\cite{EthereumDocsPlasma} or Rollups~\cite{EthereumDocsRollups}, the PureLottery protocol could play a crucial role. Rollups are solutions that execute transactions outside the main chain (Layer 1) but post transaction data on it, effectively enhancing the network's throughput. 
There are two kinds of rollups: Optimistic and zero-knowledge (ZK) rollups. In ZK rollups, the rollup operator or aggregator computes the ZK proof for all transactions on one batch and submits them to the main chain~\cite{9983760}. It is possible to randomly select the rollup aggregator for every batch to increase decentralization. By employing the PureLottery protocol, these Layer 2 solutions can ensure that the nodes responsible for these tasks are chosen in a manner that is not only fair and random but also transparent and verifiable, thereby bolstering trust and reliability in these scaling solutions.

\subsection{Protocols in Distributed Computing}
Distributed systems involve multiple computers working together on a task, offering scalability and resource sharing. However, they face challenges like data consistency and system failure management. Leader election is key in these systems for achieving consensus and coordinating tasks, ensuring smooth operation without compromising security or performance.

Peer-to-peer networks, a type of distributed system, facilitate direct communication and resource sharing among peers. Randomized leader election algorithms are vital in these networks for processes like distributed match-making and network size estimation, requiring fair and efficient decision-making in a trust-limited environment~\cite{Ramanathan2007}.

\paragraph{Data Transmission Systems and Routing.}
In data transmission systems and routing networks, the elected leader or coordinator plays a central role, often requiring frequent interaction with other nodes. For instance, in video conferencing systems or multiplayer games, the coordinator manages the status of participants and redistributes data. Leader election is thus integral to the efficient functioning of these distributed applications, ensuring effective communication and management within the network~\cite{4196222}.

\paragraph{Gossip Protocols.}
Gossip protocols are used in distributed systems for disseminating information across the network. In some implementations, a leader election process is used to select nodes that have special roles, like aggregating data or initiating certain processes~\cite{6688727}.

\section{Conclusion}
In this thesis, we introduced PureLottery, a new decentralized method for choosing leaders that does not need a separate source of random numbers. A key idea in our research is recognizing that explicit RNG is not essential for effective LE. This crucial difference between LE and RNG was not adequately addressed in prior blockchain-based LE frameworks. In PureLottery, the winner is picked using a process like a knockout tournament in sports involving multiple rounds. We showed that PureLottery fairly chooses a winner uniformly at random if everyone plays honestly, and it encourages honest play. It is also highly resistant to bias, preventing any group from unfairly increasing their winning chances or stopping the process from harming an honest participant's chances. PureLottery is efficient in its use of resources and requires only four short messages from each participant on average.

Our research further elaborates on the adaptability of the PureLottery protocol to a variety of scenarios beyond mere leader election. This includes efficient player ranking, simultaneous elections of multiple leaders, and scenarios where leader selection may be undesirable. These adaptations underscore PureLottery's versatility, showcasing its capacity to maintain security and fairness across diverse applications with minimal communication overhead.

The protocol's implementation is now openly available in the public domain. PureLottery's integration potential with smart contracts, its applicability to financial audits, public funding, and DAOs, as well as its enhancement of consensus mechanisms in blockchain technologies, demonstrate its broad applicability and potential for further research.

\newpage

\bibliography{references}
\bibliographystyle{plain}

\noindent In this work, I used ChatGPT 4 by OpenAI (chat.openai.com, accessed between November 2023 and February 2024) to help with text formulations (for example, using prompts such as "Rewrite the following text. Stay concise") and to generate ideas for further aspects in Sections~\ref{sec:intro} and \ref{sec:applications} (with prompts like "list network protocols that utilize randomized leader election").

\end{document}